\documentclass[twocolumn,preprintnumbers,floatfix,amsmath,a4paper,amssymb,nofootinbib, showpacs]{revtex4}

\usepackage{amsmath}
\usepackage{graphicx}
\usepackage{dcolumn}
\usepackage{bm}
\usepackage{amssymb}
\usepackage{epstopdf}
\usepackage{color}

\begin{document}

\title{Extreme robustness of scaling in sample space reducing processes explains Zipf's law in diffusion on directed networks}{}

\author{Bernat Corominas-Murtra$^{1}$, Rudolf Hanel$^{1}$ and Stefan Thurner$^{1,2,3}$
}

\affiliation{
$^1$ Section for the Science of Complex Systems; CeMSIIS; Medical University of Vienna; 
Spitalgasse 23; A-1090; Vienna, Austria\\
$^2$ Santa Fe Institute; 1399 Hyde Park Road; Santa Fe; NM 87501; USA\\
$^3$ IIASA, Schlossplatz 1, 2361 Laxenburg, Austria
}

\date{Version \today}

\begin{abstract}
It has been shown recently that a specific class of path-dependent stochastic processes, which reduce their sample space as they unfold,
lead to exact scaling laws in frequency and rank distributions. Such Sample Space Reducing processes (SSRP)  offer an alternative 
new mechanism to understand the emergence of scaling in countless processes. 
The corresponding power law exponents were shown to be related to noise levels in the process. 
Here we show that the emergence of scaling is not limited to the simplest SSRPs, but holds  
for a huge domain of stochastic processes that are characterized by non-uniform prior distributions. We demonstrate mathematically 
that in the absence of noise the scaling exponents converge to $-1$ (Zipf's law) for almost all prior distributions. 
As a consequence it becomes possible to fully understand targeted diffusion on weighted directed networks 
and its associated scaling laws in node visit distributions. The presence of cycles can be properly interpreted as playing the same role as noise in SSRPs and,  accordingly, determine the scaling exponents.
The result that Zipf's law emerges as a generic feature of diffusion on networks, 
regardless of its details, and that the exponent of visiting times is related to the amount of 
cycles in a network could be relevant for a series of applications in traffic-, transport- and supply chain management.
\end{abstract}


\maketitle


\section{Introduction}
Many stochastic processes, natural or man-made, are explicitly path-dependent. Famous examples include 
biological evolution \cite{Gould:1989,Kauffman:1993,Szathmary:1994} or technological innovation \cite{Arthur:1994, Sole:2013}.
Formally, path-dependence means that the probabilities to reach certain states of the system 
(or the transition rates from one state to another) at a given time depend on the history of the process up to this time. 
This statistical time-dependence can induce dramatic deformations of phase-space, in the sense that certain regions 
will hardly be revisited again, while others will be visited much more frequently.
This makes a large number of path-dependent complex systems, and processes that are associated with them, non-ergodic. 
They are typically mathematically intractable with a few famous exceptions, including the Pitman-Yor or `Chinese Restaurant' process 
\cite{Pitman:1997, Pitman:2006}, recurrent random sequences proposed by S. Ulam and M. Kac \cite{Beyer:1969,Kac:1970, Clifford:2008}, 
P\'olya urns \cite{Polya:1923, Pitman:2006, Hanel:2015}, and the recently introduced sample space reducing processes (SSRPs)  \cite{Corominas-Murtra:2014a}. 

\begin{figure*}
\includegraphics[width= 18.0cm]{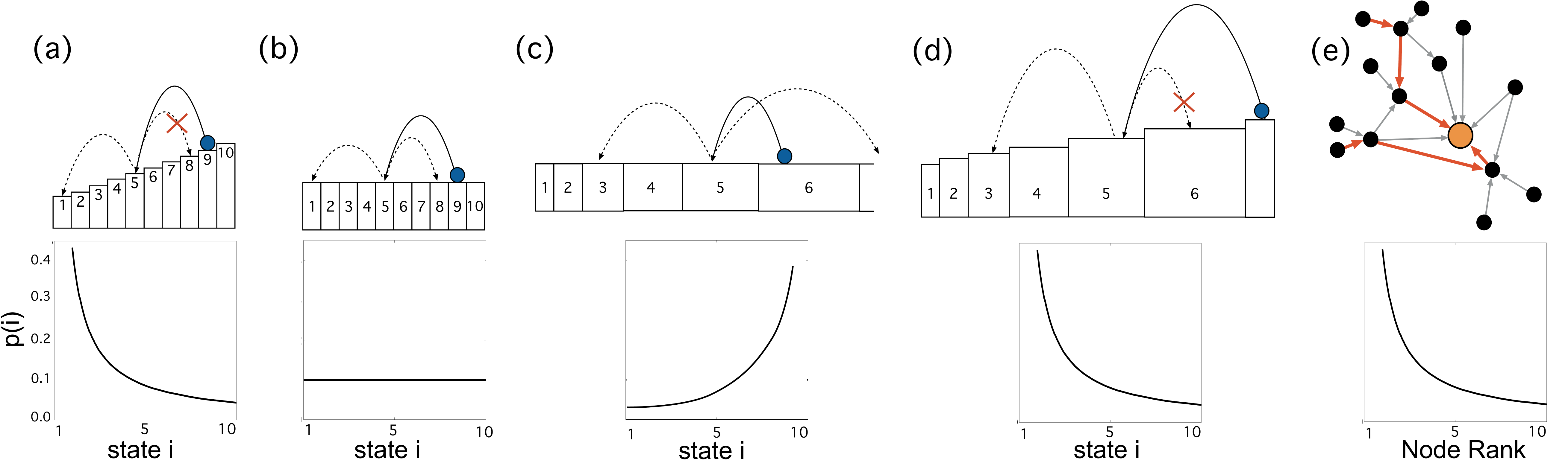}
\caption{
(a) Pictorial view of a SSRP with uniform priors. A ball  bounces downwards only with random step sizes. After many iterations of the process, the 
visiting probabilities of states $i$ approach $\sim i^{-1}$ (Zipf's law).
(b) Random process where a ball bounces random distances to the left  or right over equally-sized boxes (uniform priors). Visiting probabilities $p(i)$ are uniform.
(c) Random process as in (b) but with non-uniform prior probabilities of states (width of boxes). The visiting probabilities follow the prior probabilities.
(d) SSRP with non-uniform prior probabilities. Visiting distributions follow the attractor to a Zipf's distribution. This is true for a wide class of prior probabilities.
(e) SSRP realized by a diffusion process on a directed acyclic network towards a target node (orange). 
The visiting probability of nodes follows a Zipf's distribution, independent of the network topology.
}
\label{fig1}
\end{figure*}

SSRPs are processes that reduce their sample space as they progress over time. In their simplest form they can be depicted by the following process. Imagine a staircase like the one shown in figure \ref{fig1}a. Each state $i$ of the 
system corresponds to one particular stair. A ball is initially ($t=0$) placed at the topmost stair $N$, and can jump randomly to any of the $N-1$ lower stairs
in the next timestep with a probability $1/(N-1)$. Assume that at time $t=1$ the ball landed at  stair $i$. Since it can only jump to stairs $i'$ that are 
below $i$, the probability to jump to  stair $i'<i$ is $1/(i-1)$. The process continues until eventually stair $1$ is reached; it then halts. 

Remarkably, the statistics over a large number of repetitions of SSRPs yields an exact Zipf's law in the rank-frequency 
distribution of the visits of states \cite{Corominas-Murtra:2014a}, 
a fact that links path-dependence with scaling phenomena in an intuitive way. 
SSRPs add an alternative and independent route to understand the origin of scaling (Zipf's law in particular) to the well known
classical ways \cite{Newman:2005,Mietzenmacher:2003}, 
criticality \cite{Stanley:1987}, 
self-organised criticality \cite{Bak:1987, Jensen:1996}, 
multiplicative processes with constraints \cite{Gabaix:1999, Saichev:2009, Malevergne:2013},  
and preferential attachment models \cite{Yule:1925, Simon:1955}. 
Beyond their transparent mathematical tractability, SSRPs seem to have a wide applicability, including 
diffusion on complete directed acyclical graphs \cite{Corominas-Murtra:2014a}, 
quantitative linguistics  \cite{Corominas-Murtra:2014b}, 
record statistics \cite{Nevzorov:2001, Jensen:2013}, 
and fragmentation processes \cite{Krapivsky:1994}.

SSRPs can be seen as very specific non-standard sampling processes, with a directional bias 
or a symmetry breaking mechanism. 
In the same pictorial view as above a standard sampling processes can be depicted as a ball bouncing randomly left and right 
(without a directional bias as in the SSRP) over a set of states, see figure \ref{fig1}b. 
The ball samples the states with a uniform prior probability, meaning that all states 
are sampled with equal probability. A situation with non-uniform priors is shown in figure \ref{fig1}c
where the different widths of boxes represent the probability to hit a particular state. In a standard sampling process  
exactly this non-uniform prior distribution will be recovered.

So far, SSRPs have been studied for the simplest case only, where the potential outcomes or states 
are sampled from an underlying uniform prior distribution  \cite{Corominas-Murtra:2014a}. 
In this paper we demonstrate that a much wider class of SSRPs leads to exact scaling laws. 
In particular we will show that SSRPs lead to Zipf's law irrespective of the underlying prior distributions. 
This is schematically shown in figure \ref{fig1}d, where the prior distribution is non-uniform, and states are sampled 
with a SSRP. The resulting distribution function will no longer follow the prior distribution as in figure \ref{fig1}c, 
but produces Zipf's law. 
We provide show in detail  how SSRPs depend on their prior distributions. Zipf's law turns out to be 
an attractor distribution that holds for practically any SSRP, irrespective of the details of the stochastic system at hand, 
i.e. irrespective of their prior distributions. 
This extreme robustness with respect to details of transition rates between states within a system 
offers a simple understanding of the ubiquity of  Zipf's law.  Phenomena that show a high robustness of 
Zipf's law with respect to changes on the detailed properties of the system have been reported before 
\cite{Nevzorov:2001, Jensen:2013, Corominas-Murtra:2010}. 

As an important example we demonstrate these mathematical facts in the context of diffusion processes 
on {\em Directed Acyclic Graphs} (DAG). Here  Zipf's distributions of node visiting frequencies 
appear generically, regardless of the weight- or degree distribution of the network. 
We call diffusion processes 
on DAG structures {\em targeted diffusion}, since, in this type network, diffusion is targeted towards a set of target or sink nodes, see figure \ref{fig1}e. 
The targeted diffusion results we present here are in line with recent findings reported in \cite{Perkins:2014}.
 
\section{SSRPs with arbitrary priors}
We start the formal study of the statistics of SSRPs for the noiseless case which implies -- in the staircase picture -- 
that upward jumps are not allowed (sampling with a bias). We then study how the statistics of SSRPs behaves 
when noise is introduced. In this case the probability of upward jumps is no longer zero. 

\subsection{Noiseless SSRPs}
Think of the $N$ possible states of a given system as stairs with different widths and imagine a ball bouncing 
downstairs with random step sizes. The probability of the downward bouncing ball to hit stair $i$ is proportional 
to its width $q(i)$, see figure \ref{fig1}d. Given these prior probabilities $q(i)$, the transition probability from 
stair $j$ to  stair $i$ is 
\begin{equation}
p(i|j)=\left\{\begin{array}{ll}
\frac{q(i)}{g(j-1)}\;\;{\rm if}\;i<j \\
0\;{\rm otherwise},
\end{array}
\right.
\label{eq:P(a_k|a_i)}
\end{equation}
with $g(j-1)=\sum_{\ell<j}q(\ell)$. Prior probabilities  are normalised, $\sum_i q(i)=1$. We denote such a SSRP by $\psi$.
One can safely assume the existence of a stationary visiting distribution, $p$, arising from many repetitions of process $\psi$ and satisfying the following relation:
\begin{eqnarray}
p(i)=\sum_{i<j\leq N}p(i|j)p(j) \quad .
\label{SIeq:Consistent}
\end{eqnarray}
Using equation  (\ref{eq:P(a_k|a_i)}), and forming the difference 
\begin{equation}
\frac{p(i+1)}{q(i+1)}-\frac{p(i)}{q(i)}=-\frac{p(i+1)}{g(i)} \quad , 
\end{equation}
and by re-arranging terms we find that 
\begin{equation}
\frac {p(i+1) g(i+1)}{q(i+1)}=\frac{p(i) g(i)}{q(i)} \quad , 
\end{equation}
where we use the fact that $g(i)+q(i+1)=g(i+1)$. 
Note that this is true for all values of $i$, and  in particular 
\begin{equation}
\frac {p(i) g(i)}{q(i)}=\frac{p(1) g(1)}{q(1)}  =  p(1)\quad , 
\end{equation}
since $g(1)=q(1)$. We arrive at the final result 
\begin{equation}
p(i)=\frac{q(i)}{g(i)}p(1) \qquad {\rm with} \qquad    \frac{1}{p(1)}=\sum_{j\leq N}\frac{q(j)}{g(j)}  \quad .
\label{eq:SolutionNo_Noise} 
\end{equation}
$p(i)$ is the probability that we observe the ball ball bouncing downwards at stair $i$.
Equation (\ref{eq:SolutionNo_Noise}) shows that the  path-dependence of the SSRP $\psi$  {\em deforms} the prior 
probabilities of the states of a given system, 
$q(i)\to p(i)= \frac{q(i)}{g(i)}$.
We can now discuss various concrete prior distributions. 
Note that equation (\ref{eq:SolutionNo_Noise}) is exact and does not dependent on system size. 
\\

\begin{figure}[t]
\includegraphics[width= 8.7 cm]{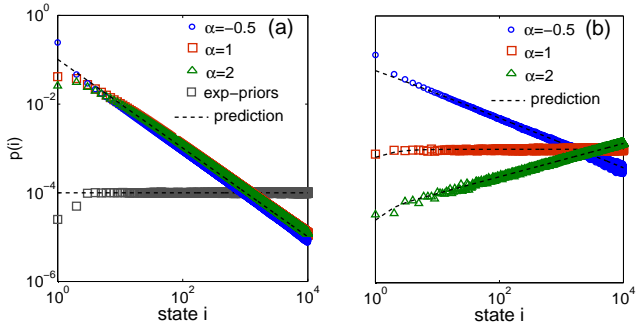}
\caption{
Probability distributions arising from numerical realizations of  SSRPs over $10^4$ states without noise (a), and with a noise level of $\lambda=0.5$, (b).
Colors correspond to various prior probabilities: polynomial, $q(i)\sim i^{\alpha}$, with $\alpha=-0.5$ (blue circles), $\alpha=1$ (red circles) and $\alpha=2$ (green circles) in both panels.  The exponential case, $q(i)\sim e^i$ (grey squares) is shown in panel (a) only. 
Dashed black lines show the theoretical results without noise from equation (\ref{eq:SolutionNo_Noise}) (a), and with noise from equation  (\ref{eq:PowerPriorsNoise}) (b). 
Clearly, Zipf's law ($p(i)\sim i^{-1}$) emerges for the different polynomial prior probabilities, whereas for the exponential prior probability 
the expected uniform distribution is obtained (a). All simulations were done with $10^7$ repetitions (a) and $10^5$ repetitions (b). 
 }
\label{fig:Numerics}
\end{figure}

{\em Polynomial priors and the ubiquity of Zipf's law:} 
Given power law priors, $q(i) \sim i^{\alpha}$ with $\alpha>-1$, one can compute $g$ up to a normalisation constant
\begin{equation}
g(i)=\sum_{j\leq i}{j^\alpha}=\frac{i^{\alpha+1}}{\alpha+1}+{\cal O}(i^{\alpha}) \quad , 
\end{equation}
which, when used in equation (\ref{eq:SolutionNo_Noise}), asymptotically gives
\begin{eqnarray}
p(i) \sim \frac{p(1)}{i}\quad,
\label{eq:TranformationZipfPol}
\end{eqnarray}
i.e., Zipf's law. More generally, this result is true for polynomial priors, $q(j)\sim\sum_{i\leq m}a_i j^{\alpha(i)}$, where the degree of the polynomial 
$\alpha(m)=\max\{\alpha(i)\}$ is larger than $-1$, in the limit of large systems. Numerical simulations show perfect agreement with the theoretical prediction for various 
values of $\alpha$, see figure \ref{fig:Numerics}a (circles, triangles, red squares). 
\\

{\em Fast decaying priors:}
The situation changes drastically for exponents $\alpha <-1$.
For sufficiently fast decaying priors we have  
\begin{equation}
g(i) \sim \int_1^i q(x)dx \sim  g(1)=q(1)\quad .
\end{equation}
The fast decay makes the contribution to $g$ from large $i$'s negligible.
Under these circumstances equation (\ref{eq:SolutionNo_Noise}) can be approximated for sufficiently large $i$'s, as $p(i) \sim q(i)$.  
We encounter the remarkable situation that for fast decaying priors the SSRP, even though it is history dependent, 
follows the prior distribution. In this case the SSRP resembles a standard sampling process.
\\

{\em Exponential priors:} 
For exponential priors, $q(i)\sim e^{\beta i}$, with $\beta>0$, we find according to equation  (\ref{eq:SolutionNo_Noise})   
that $p(i)=1/N $, i.e., a uniform distribution. To see this note that, up to a normalisation constant, $g(i)$ is a geometric series,
\begin{eqnarray}
g(i)&=&\sum_{j=1}^ie^{\beta j} 
=e^\beta\frac{e^{\beta i}-1}{e^{\beta} -1}\quad. \nonumber
\end{eqnarray}
Substituting it into equation (\ref{eq:SolutionNo_Noise}), one finds the exact relation
\begin{equation}
p(i)=p(1)\frac{1-e^{-\beta}}{1-e^{-\beta i}}\quad,
\end{equation}
which can be safely approximated,  for $i\gg 1$, by
\begin{equation}
p(i)\to p(1)\left(1-\frac{1}{e^\beta}\right)\quad.
\end{equation}
We observe that this is a constant independent of $i$. Accordingly, after normalisation, we will have $p(i) \sim 1/N$.
Note that exponential priors describe a somewhat pathological situation. Given that a state $i$ is occupied 
at time $t$, the probability to visit state $i-1$ is huge compared to all the other remaining states, 
so that practically all states will be sampled in a descending sequence: $i \to i-1\to i-2\to i-3\to \cdots 1$, 
which obviously leads to a uniform $p$.
Again, numerical simulations show perfect agreement with the prediction, as shown in  figure \ref{fig:Numerics}a (grey squares). 
Switching from polynomial  to exponential priors, we switch the attractor from the Zipf's regime to the uniform distribution. 

\subsection{Noisy SSRPs}
Noisy SSRPs are mixtures of a SSRP $\psi$ and stochastic transitions between 
states that are not history-dependent. Following the previous scheme of the staircase picture,  
the {\em noisy} variant of the SSRP, denoted by  $\psi_\lambda$, starts at 
$N$ and jumps to any stair $i<N$, according to the prior probabilities $q(i)$. At $i$ the process now has two options: 
(i) with probability $\lambda$ the process  continues the SSRP and jumps to any $j<i$, or, 
(ii) with probability $1-\lambda$ jumps to {\em any} point $j<N$, following a standard process of sampling without memory. 
$1-\lambda$ is the noise strength.
The process stops when stair $1$ is hit. The transition probabilities for $\psi_\lambda$ read, 
\begin{equation}
p(i|j)=\left\{\begin{array}{ll}
\lambda\frac{q(i)}{g(j-1)}+(1-\lambda)q(i)\;\;{\rm if}\;i<j \\
(1-\lambda)q(i)\;{\rm otherwise}\quad.
\end{array}
\right.
\label{eq:P(a_k|a_i)lambda}
\end{equation}
Note that the noise allows moves from $j$ to $i$, even if $i>j$. 
Proceeding exactly as before we get
\begin{equation}
\frac{p_\lambda(i+1)}{q(i+1)}\left(1+\lambda\frac{q(i+1)}{g(i)}\right)=\frac{p_\lambda(i)}{q(i)} \quad ,
\end{equation}
where $p_\lambda(i)$ depicts the probability to visit state $i$ in a noisy SSRP with parameter $\lambda$.
As a consequence we obtain:
\begin{equation}
p_\lambda(i)=p_\lambda(1)\frac{q(i)}{q(1)}\prod_{1<j\leq i}\left(1+\lambda\frac{q(j)}{g(j-1)}\right)^{-1}\,.
\label{eq:Noise_exact}
\end{equation}
The product term can be safely approximated by
\begin{eqnarray}
\prod_{1<j\leq i}\left(\cdots\right)^{-1}
&=&\exp\left[-\sum_{1<j\leq i}\log\left(1+\lambda\frac{q(j)}{g(j-1)}\right)\right]\nonumber\\
&\approx&\exp\left[-\sum_{1<j\leq i}\lambda\frac{q(j)}{g(j-1)}\right]\nonumber\\
&\approx&\exp\left[-\lambda\log\left(\frac{g(i)}{q(1)}\right)\right]\nonumber\\
&=&\left(\frac{g(i)}{q(1)}\right)^{-\lambda}\quad ,
\end{eqnarray}
where we used $q(j) \sim  \left. dg/dx\right|_{j}$ and $\log(1+x) \sim  x$ for small $x$, assuming that  $x=\lambda\frac{q(j)}{g(j-1)}\ll 1$.
Finally, we get 
\begin{equation}
p_\lambda(i) \sim \frac{p_\lambda(1)}{q(1)^{1-\lambda}}\left(\frac{q(i)}{g(i)^\lambda}\right)\quad,
\label{eq:General_Noise}
\end{equation}
where $p_\lambda(1)/q(1)^{1-\lambda}$ acts as the normalisation constant. 
$\lambda$ plays the role of a scaling exponent. 
For $\lambda\to 1$ (no noise), $p_\lambda$ recovers the standard SSRP $\psi$ of equation  (\ref{eq:P(a_k|a_i)}). 
For $\lambda=0$, we recover the case of standard random sampling, $p \to q$. It is worth noting that continuous SSRP 
display the same scaling behaviour (see Appendix A).
The particular case of $q(i)=1/N$ that was  studied in \cite{Corominas-Murtra:2014a}, shows that $\lambda$  
turns out to be the scaling exponent of the distribution $p_\lambda(i)\sim1/i^{\lambda}$.
Note that these are not frequency- but rank distributions. They are related, however. The range of exponents 
$\lambda\in (0,1]$ in rank, represents the respective range of exponents $\alpha\in [2,\infty)$ in frequency, see e.g. 
\cite{Newman:2005} and Appendix B. For polynomial priors, $q(i)\sim i^{\alpha}$  ($\alpha>-1$), one finds
\begin{equation}
p_\lambda(i)\sim i^{\alpha(1-\lambda)-\lambda}\quad.
\label{eq:PowerPriorsNoise}
\end{equation}
The excellent agreement of these predictions with numerical experiments is shown in figure \ref{fig:Numerics}b.
Finally, for exponential priors $q(i)\sim e^{\beta i}$  ($\beta>0$) the visiting probability of for the noisy SSRP $\psi_\lambda$ becomes
$p(i)\sim e^{(1-\lambda)\beta i}$, see Tab. \ref{tab}. 
Clearly, the presence of noise recovers the prior probabilities in a fuzzy way, depending on the noise levels. 
The following table sumarizes the various scenarios for the  distribution functions $p(i)$ for the different 
prior distributions $q(i)$ and noise levels. 
\begin{table}[ht!]
\caption{Distribution functions $p(i)$ of SSRPs for the various prior distributions $q(i)$. SSRP distributions with a noise level of 
$(1-\lambda)$ are indicated by $p_\lambda(i)$. 
}
 \begin{tabular}{l c c c } 
 \hline
 \hline
prior &  (sub-) logarithmic & polynomial   & exponential\\
$q(i)$ & $ i^{\alpha}$ ($\alpha <-1$) & $i^{\alpha}$ ($\alpha>-1$) & $e^{\beta i}$\\
 \hline
$p(i)$ &  $i^{\alpha}$ & $ i^{-1}$ & $\frac1N$ \\ 
$p_\lambda(i)$ noise & $ i^{\alpha}$ & $  i^{\alpha(1-\lambda)-\lambda}$  & $ e^{(1-\lambda)\beta i} $  \\ 
 \hline
  \hline
\end{tabular}
\label{tab}
\end{table}

\section{Diffusion on weighted, directed, acyclic graphs}
\label{sec3}

The above results have immediate and remarkable consequences for the diffusion on
DAGs \cite{Karrer:2009} or, more generally, on networks with target-, sink- or absorbing nodes.
We call this process {\em targeted diffusion}.
In particular, the results derived above allow us to understand the origin of Zipf's law of node visiting
times for practically all weighted DAGs, regardless of their degree-  and  weight distributions.
We first demonstrate this fact with simulation experiments on weighted DAGs and then,
in section \ref{sec:AnalyticsRandomDAG} we analytically derive the corresponding equations of targeted
diffusion for the large class of sparse random DAGs, that explain that Zipf's law must occur in node visiting frequencies.
In  appendix B proofs are given for the cases of exponential and scale free networks.

\begin{figure}
\includegraphics[width= 8.5cm]{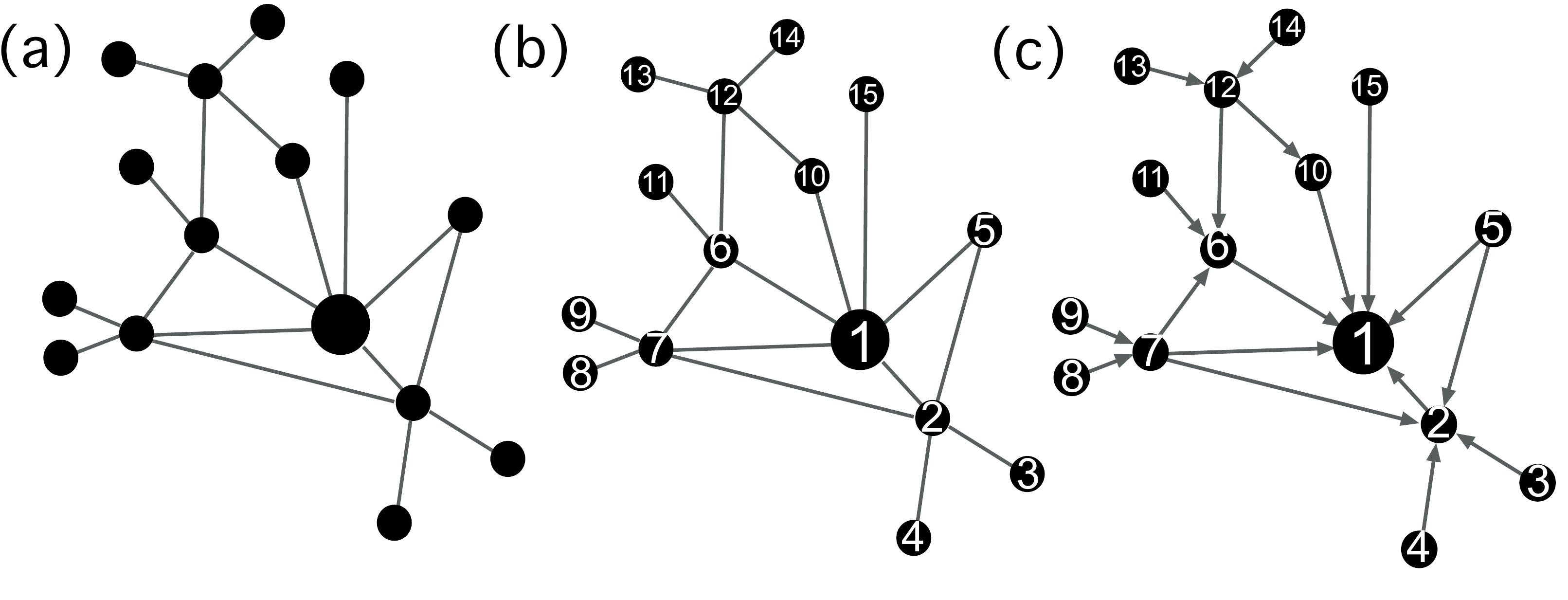}
\caption{
Building a DAG. (a) Start with any undirected, connected graph. (b) Place a unique label $1, . . .,N$ on each node of the graph. (c)
draw an arrow from $i$ to $j$, if $i>j$, or  from $j$ to $i$,  if $i<j$. The strict ordering induced by the labelling prevents the emergence
of cycles \cite{Karrer:2009, Goni:2010}. Such a graph will have at least, one {\em target} or a {\em sink} node, in the depicted case this is node $i=1$.
A diffusion process of this graph, where random walkers are randomly placed on the graph and follow the arrows at every timestep,
is called {\em targeted diffusion} with target node $i=1$.}
\label{fig:DAGS}
\end{figure}

We start with the observation that SSRPs with uniform priors can be seen as a diffusion processes
on a fully connected DAG, where nodes correspond one-to-one to the stairs of the above examples.
This results in a Zipf's law of node visiting frequencies \cite{Corominas-Murtra:2014a}.
However, such fully connected networks are extremely unlikely to occur in reality.
To create much more realistic structures, we generate arbitrary random DAGs following e.g. references \cite{Karrer:2009, Goni:2010}.
Start with any undirected connected graph ${\cal G}(V, E)$, with $V$  the set of nodes, $E$ the set of edges, and $P(k)$ the degree distribution, see
figure \ref{fig:DAGS}a.
Next, label each node in any desired way that allows an ordering, for example with numbers $1, . . .,N$, see figure \ref{fig:DAGS}b.
The labelling induces an order that determines the directionality of links in the graph: if nodes $i$ and $j$ are connected,
we draw an arrow from $i$ to $j$, if  $i>j$, or from $j$ to $i$, if $i<j$, as seen in figure \ref{fig:DAGS}c.
We denote the resulting DAG by  ${\cal G}^D(V, E^D)$.
The order induced by the labelling mimics the order (or symmetry breaking) that underlies any SSRPs. By definition, there exists, at least, one target node, "$1$".

Noise can be introduced to this DAG construction as follows: if node $i$ and $j$
are connected in ${\cal G}$ and $i>j$ one can assign an arrow from $i$ to $j$ (as before) with probability $\lambda$, or
place the arrow in a random direction with probability $1-\lambda$. This will create cycles
that play the role of  {\em noise} in the targeted diffusion process. This network is no longer a pure DAG since it contains cycles.

\subsection{Targeted diffusion on specific networks}

\begin{figure}
\includegraphics[width= 8.9cm]{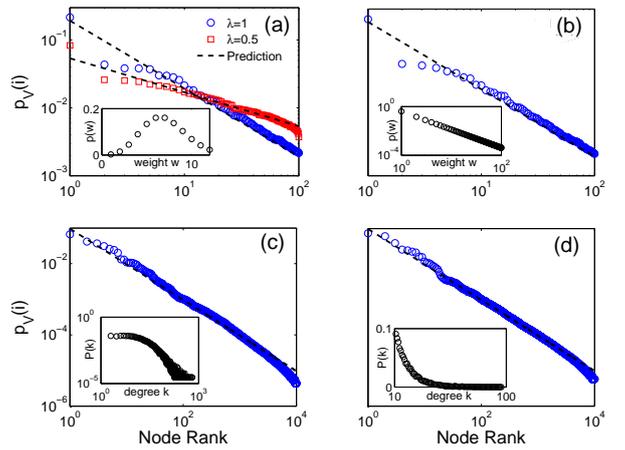}
\caption{
Node visiting rank distributions from diffusion on weighted DAGs, built over Erd\H{o}s-R\'enyi graphs (see DAG construction)
with $p=1/2$, and $N=100$ nodes (a) and (b).
The weight distribution $w_{ik}$ follows (a) a Poisson distribution with average $\mu=6$, and (b) a power-law $p(w)\propto w^{-1.5}$
that is shown in the inset.
In both cases the predicted Zipf's law is present (black dashed line), even though the networks are small. In (a) the DAG condition is
violated (red squares) by  assigning random directions to a fraction of $1-\lambda$ links. This allows for the presence of cycles,
which play the role of noise in a SSRP. A power law with the exponent $\lambda$ is observed in the corresponding
rank distribution, perfectly in line with the theoretical predictions (dashed black lines).
(c) A targeted diffusion experiment on a DAG that is based on the citation network of HEP ArXiv repository, containing $10^4$ nodes belonging to the $10^4$ most cited papers.
(d) The results of the same experiment on an exponential network of the same size is given. The inset shows the respective degree distributions. Despite the huge topological difference between these two graphs, the rank distribution of visits to nodes is clearly of Zipf's type
for almost four decades in both cases.
}
\label{fig3}
\end{figure}

A diffusion process on ${\cal G}^D$ is now carried out by placing random walkers on the nodes randomly, and letting them
take steps following the arrows in the network. They diffuse according to the weights in the network until they hit a target node and are then removed.
We record the number of visits to all nodes and sort them according to the number of visits,
obtaining a {\em rank distribution} of visits\footnote{Rank ordering is not necessary whatsoever to see the clear agreement
with the theoretical predictions. Almost identical results are seen when we order nodes according to their numerical ordering.}.
We show the results from numerical experiments of $10^7$ random walkers on various DAGs in figure \ref{fig3}.
In Figs. \ref{fig3}a and \ref{fig3}b we plot the rank distribution of visits to nodes for weighted Erd\H{o}s-R\'enyi (ER) DAG networks.
A weight $w_{ik}$  is randomly assigned to each link $e_{ik}\in E$ from a given weight distribution $p(w)$. Weights either follow a Poisson distribution, figure \ref{fig3}a,
or a power-law distribution, figure \ref{fig3}b. In both cases Zipf's law is obtained in the rank distribution of node visits.
For the same network we introduce {\em noise} with $\lambda=0.5$  and carry out the same diffusion experiment.
The observed slope corresponds nicely with the predicted value of $\lambda$, as shown in figure \ref{fig3}a (red squares) for the Poisson weights.

We computed rank distributions of node visits from diffusion on more general network topologies.
In figure \ref{fig3}c we show the rank distribution of node visits where the substrate network is the citation network
of High Energy Physics  in the ArXiv repository \cite{HEPNET, Leskovec:2007}, and the order is induced by the degree of nodes.
Figure \ref{fig3}d shows the rank distribution of node visits from diffusion on an exponential DAG, that is generated by non-preferential attachment \cite{Callaway:2001}, where the order of nodes is again induced according to the degree. Both networks show Zipf's law in the rank distribution of node visits.
This is remarkable since both networks are drastically different in topological terms.

\subsection{Analytical results for targeted diffusion on random DAGs}
\label{sec:AnalyticsRandomDAG}

For diffusion on random DAGs it is possible to obtain analytic results that are identical to equation (\ref{eq:P(a_k|a_i)}), showing that
Zipf's law is generally present in targeted diffusion.

We first focus on the definition of the prior probabilities in the context of diffusion on undirected networks.
As stated above, $q(i)$ is the probability that state $i$ is visited in a random sampling process, see  Figs. 1b and 1c.
In the network context this corresponds to the probability that node $i$ is visited by a random walker.
Assume that we have an undirected random graph ${\cal G}(V,E)$ and that the $N$ nodes are labelled $1, . . .N$.
The probability that a random walker arrives at node $i$ from a randomly chosen link of $E$, the {\em network-prior} probability of node $i$,
is easily identified as
\begin{equation}
q_G(i)\equiv\frac{k_i}{2|E|}\quad,
\label{eq:PriorNets}
\end{equation}
where $|E|$ is the number of links in the graph; the factor $2$ appears because a link contains $2$ endpoints.
If $\sigma_G \equiv \{k_1,. . .,k_N\}$ denotes the undirected degree sequence $q_G$, is a simple rescaling of $\sigma_G$, i.e., $q_G=\frac{1}{2|E|}\sigma_G$.
Using the same notation as before, the cumulative {\em network-prior} probability distribution is $g_G(i)\equiv\sum_{\ell \leq  i}q_G(\ell)$.

From equation (\ref{eq:PriorNets}) and by assuming that in sparse graphs the probability of self-loops vanishes, i.e., $p(e_{ii})\to 0$,
one can compute the probability that a link $e_{ij}$ exists in ${\cal G}$, \cite{Chung:2002}
\begin{eqnarray}
p(e_{ij}\in E)&=& \frac{k(i)k(j)}{\sum_{\ell\leq N}k(\ell)}  
=2|E|q_G(i)q_G(j) \quad ,
\label{eq:p_eij}
\end{eqnarray}
where the second step is possible since $\sum_{\ell\leq N}k(\ell)=2|E|$. With this result,
the {\em out-degree } of node labelled $i$ in the graph ${\cal G}^D$ can be approximated by
\begin{eqnarray}
k^{\rm out}_i&=&\sum_{j<i}p(e_{ij}\in E)\nonumber\\
&=&2|E|\sum_{j<i}q_G(i)q_G(j)\nonumber\\
&=&2|E|q_G(i)\sum_{j<i}q_G(j)  \nonumber\\
&=&  2|E|q_G(i) g_G(i-1)   \quad .
\label{eq:kout}
\end{eqnarray}
Note that to compute $k^{\rm out}_i$ we only need take into account the 
(undirected) links which connect $i$ to nodes with a lower label $j<i$, according to the labelling used for the DAG construction outlined above.

We can now compute the probability that a random walker jumps from node $i$ to node $j$ on the DAG ${\cal G}^D$,
\begin{equation}
p_G(j|i)=\left\{
\begin{array}{ll}
p(j|i, e_{ij}\in E)p(e_{ij}\in E)\;\;{\rm if}\;i>j\\
0\;\;{\rm otherwise} \quad .
\end{array}
\right.
\label{eq:p(k|i)General}
\end{equation}
This is the network analogue of equation (\ref{eq:P(a_k|a_i)}).
Here $p(j|i, e_{ij}\in E)$ is the probability that the random walker jumps from $i$ to $j$ given that $i>j$ and the link $e_{ij}$ exists in ${\cal G}$. Clearly,
this probability is
\begin{eqnarray}
p(j|i, e_{ij}\in E)&=&\frac{1}{k^{\rm out}_i}     \nonumber\\
&=&\left(2|E|q_G(i)   g_G(i-1)\right)^{-1}  ,
\label{eq:kout2}
\end{eqnarray}
Using Eqs. (\ref{eq:p_eij}) and (\ref{eq:kout2}) in equation (\ref{eq:p(k|i)General}) we get
\begin{equation}
p_G(j|i)=\left\{
\begin{array}{ll}
\frac{q_G(j)}{g_G(i-1)};\;{\rm if}\;i>j\\
0\;\;{\rm otherwise}\quad,
\end{array}
\right.
\label{eq:q-factorization}
\end{equation}
which has the same form as equation (\ref{eq:P(a_k|a_i)}).
Note that this expression only depends on $q_G$, i.e. the degrees of nodes in the {\em undirected } (!)  graph ${\cal G}$.
The solution of equation (\ref{eq:q-factorization}) is obtained in exactly the same way as before for equation (\ref{eq:P(a_k|a_i)}),
and the node visiting probability of targeted diffusion on random DAGs is
\begin{equation}
p_{\rm }(i)\propto\frac{q_G(i)}{g_G(i)} \quad,
\label{eq:General_Net}
\end{equation}
which is the network analog of equation (\ref{eq:SolutionNo_Noise}).

We finally show the results for a DAG that is based on an ER graph.
For an ER graph, by definition, the probability for a link to exist is a constant $r \in(0,1]$, and $p(e_{ij}\in E)=r$.
Again we label all nodes by $1, . . .,N$ and build a DAG ${\cal G}_{ER}^D$ as described above.
It is not difficult to see that the out-degree of node $i$ is $k^{\rm out}(i)=(i-1) r$, and,
using this directly in equation (\ref{eq:p(k|i)General}), we get
\begin{equation}
p_G(j|i)=\left\{
\begin{array}{ll}
\frac{1}{i-1}\;\;{\rm if}\;i>j\\
0\;\;{\rm otherwise} \quad ,
\end{array}
\right.
\end{equation}
which is the standard equation  for a SSRP with uniform prior probabilities $q$, \cite{Corominas-Murtra:2014a}.
This means that for the ER graph $q_G(i)$ is a constant and $g_G(i)\sim i$. Using this in equation (\ref{eq:General_Net}),
we find that the node visiting probability is exactly Zipf's law, with respect to the ordering used to build the DAG,
\begin{equation}
p_{\rm }(i)\propto i^{-1}\quad.
\end{equation}
Note that this result is independent of $r$ and, therefore, of the average degree of the graph.

\section{Discussion}
We have shown that if a system, whose states are characterized by prior probabilities $q$, is sampled through a
SSRP, the corresponding sampling space gets deformed, in a way that Zipf's law emerges as a dominant attractor.
This is true for a huge class of reasonable prior probabilities, and might be the fundamental origin of the ubiquitous presence of Zipf's law in nature.
On the theoretical side we provide a direct link between non-ergodicity as it typically occurs in path-dependent processes
and power laws in corresponding statistics.
Formally, SSRPs define a microscopic dynamics that results in a deformation of the phase space.
It has been pointed out that the emergence of non-extensive properties may be related to
generic deformations of the phase space \cite{Hanel:2011, Hanel:2014, Hernando:2012}. Consequently, SSRPs
offer a entirely new playground to connect microscopic and macroscopic dynamics in non-equilibrium systems.
Our results could help to understand the astonishing resilience of some scaling patterns which are associated with
Zipf's law, such as the recent universality in body-mass scaling found in ecosystems \cite{Hatton:2015}.

We discussed one fascinating direct application of this process: the origin of scaling laws in node visit frequencies in targeted diffusion on networks.
We demonstrated both theoretically and by simulations that the immense robustness of these scaling laws in targeted diffusion
-- and Zipf's law in particular -- arises generically, regardless of its topological details, or weight distributions.
The corresponding exponents are related to the amount of cycles in a network. This finding should be relevant for a series
of applications of targeted diffusion on networks where a target
has to be found and reached, such as in traffic-, transport- or supply chain management.
We conjecture that these findings and variations will apply for search processes in general.

{\em Acknowledgments}.- 
This work was supported by the Austrian Science Fund FWF under P29252 {\em Generalized information theoretic approaches for history dependent-processes} 
and the FP7 projects LASAGNE no. 318132 and  MULTIPLEX no. 318132.

\newpage

\appendix
\section{Continuous SSRPs}
Consider the interval $\Omega=(0,N]$. 
The prior probability density $q$  is defined from a differentiable function $f:\Omega\to\mathbb{R}^+$ as
\begin{equation}
q(x)=\left\{
\begin{array}{ll}
f(x)\;{\rm if}\;x\in[1,N]\\
f(1)\;{\rm otherwise} \quad .
\end{array}
\right.
\label{SIeq:q}
\end{equation}
Since this represents a probability density
\[
\int_0^N q(x)dx=1 \quad.
\]
The region $(0,1)$ where $q(x)=f(1)$ acts as a {\em trapping region} of finite measure.
As we shall see, the particular choice of the length of such trapping region has no consequences for the global statistical patters, as long as it is finite. We will refer to this trapping region as $\Omega_1$. In addition, for any $x\in\Omega\setminus \Omega_1$ we define the interval $\Omega_x=(0,x)$, which is the sampling space from point $x$. These sampling spaces are now continuous but still can be ordered by inclusion, meaning that if $x,y\in\Omega$ and $x>y$, then $\Omega_y\subset\Omega_x$. 

\subsection{Noiseless continuous SSRPs}
\begin{figure}
\includegraphics[width= 8.3cm]{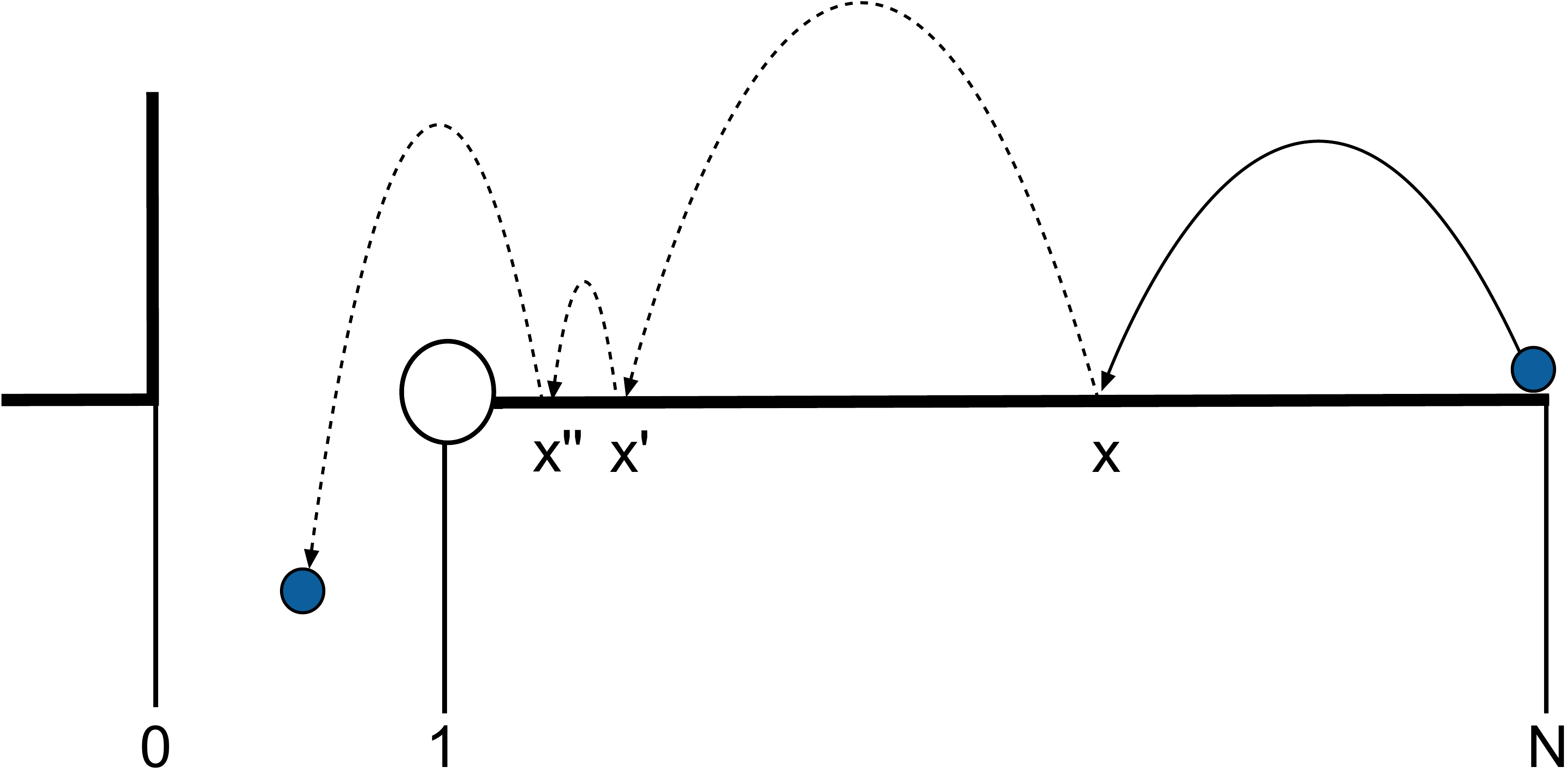}
\caption{Continuous SSRPs: A ball bouncing {\em to the left}  on a continuous interval $\Omega=[0,N]$. At each time step it 
lands at a given point of $\Omega$ according to a prior probability density $q(x)dx$. The process stops when the {\em ball} 
falls into a region of finite measure, represented here as the interval $[0,1]$. 
 }
\label{fig:Continuous}
\end{figure}

With the example of the staircase in mind, we can describe a SSRP $\psi$ over a continuous sampling space, see figure (\ref{fig:Continuous}). We start  in the extreme of the interval, $x=N$, and we choose any point of $\Omega$ following the probability density $q$. Suppose we land in $x<N$. Then, at time $t=1$ we choose at random some point $x'\in\Omega_x$ following a probability density proportional to $q$. We run the process until a point $z\in\Omega_1$ is reached. Then the process stops. The SSRP $\psi$ can be described by the transition probabilities between the elements of $x,y\in\Omega$ such that $y>1$ as follows,
\begin{equation}
p(x|y)=\left\{\begin{array}{ll}
q(x)/g(y)\;\;{\rm iff}\;x<y \\
0\;{\rm otherwise} \quad,
\end{array}
\right.
\label{SIeq:P(a_k|a_i)Cont}
\end{equation}
where $g(y)$ is the cumulative density distribution evaluated at point $y$,
\begin{equation}
g(y)=\int_{\Omega_y}q(x)dx=\int_{1}^yq(x)dx+f(1) \quad .
\end{equation}

We are interested in the probability density $p$ which governs the frequency of visits along $\Omega$ after the sampling process $\psi$. To this end, we start with the following self-consistent relation for $p$, 
\begin{eqnarray}
p(x)=\int_{x}^Np(x|y)p(y)dy \quad .
\label{SIeq:ConsistentCont}
\end{eqnarray}
Recall that the integration limits $\int^N_x$ represent the fact that a particular state $x$ can only be reached from a state $y>x$. By differentiating this integral equation we obtain:
\begin{equation}
\frac{d p}{d x}=\frac{d}{d x}\left(\int_{x}^Np(x|y)p(y)dy\right) \quad .
\label{SIeq:DirectDerivative}
\end{equation}
In agreement to equation  (\ref{SIeq:P(a_k|a_i)Cont}), $p(x|y)= q(x)/g(y)$ if $y>1$ and $y>x$.
Equation (\ref{SIeq:DirectDerivative}) can be expanded using the Leibniz rule: 
\begin{eqnarray}
\frac{d p(x)}{d x}&=&\int_{x}^N\frac{d p(x|y)}{d x}p(y)dy-\frac{q(x)}{g(x)}p(x)\nonumber\\
&=&\frac{1}{q(x)}\frac{d q(x)}{d x}\int_{x}^N\frac{q(x)}{g(y)}p(y)dy-\frac{q(x)}{g(x)}p(x)\nonumber\\
&=&\frac{1}{q(x)}\frac{d q(x)}{d x}p(x)-\frac{q(x)}{g(x)}p(x)\quad .
\end{eqnarray}
This leads to a differential equation  governing the dynamics of SSRPs under arbitrary prior probabilities $q$, 
\begin{equation}
\frac{d p(x)}{d x}=\left(\frac{1}{q(x)}\frac{d q(x)}{d x}-\frac{q(x)}{g(x)}\right)p(x) \quad .
\label{eq:Diff_eq_noNoise}
\end{equation}
The above equation  can be easily integrated in the interval $(1,N]$.
Observing that equation (\ref{eq:Diff_eq_noNoise}) can be rewritten as
\begin{equation}
\frac{dp(x)}{p(x)}=\left[\frac{d}{dx}\log\left(\frac{q(x)}{g(x)}\right)\right]dx \quad,
\end{equation}
One finds:
\begin{equation}
\log p(x)=\log\left(\frac{q(x)}{g(x)}\right)+\kappa \quad,
\end{equation}
$\kappa$ being an integration constant to be determined by normalisation.
The above equation has as a general solution for points $x\in(1,N]$
\begin{equation}
p(x)=\frac{1}{Z}\frac{q(x)}{g(x)} \quad, 
\end{equation}
where $Z$ is the normalisation constant
\begin{equation}
Z=\int_0^N\frac{q(y)}{g(y)}dy \quad.
\end{equation}
This demonstrates how the prior probabilities $q$ are {\em deformed} when sampled through the SSRP $\psi$ in the region $x\in (1,N]$. 
This is the analogous to equation (\ref{eq:SolutionNo_Noise}) of the main text.

\subsection{Continuous SSRPs with noise}
Suppose the interval $\Omega=(0,N]$ and let us define a probability density $q$ on $\Omega$ as in equation  (\ref{SIeq:q}). 
The {\em noisy} SSRP $\psi_\lambda$ starts at $x=N$ and jumps to any point in $x'\in\Omega$, according to the prior probabilities 
$q$. From $x'$ the system has two options: (i) with probability $\lambda$ the process jumps to any $x''\in \Omega_{x'}$, i.e., $\psi_\lambda$
continues the SSRP we described above or, (ii) with probability $1-\lambda$, $\psi_\lambda$ jumps to {\em any} point $x''\in\Omega$, 
following a standard sampling process. The process stops when it jumps to a member of the {\em sink} set, 
namely to a $x\leq1$. The transition probabilities now read $(\forall y>1)$, 
\begin{equation}
p(x|y)=\left\{\begin{array}{ll}
\lambda q(x)/g(y)+(1-\lambda)q(x)\;\;{\rm iff}\;x<y \\
(1-\lambda)q(x)\;{\rm otherwise} \quad,
\end{array}
\right.
\label{SIeq:P(a_k|a_i)lambdaCont}
\end{equation}
Note that the noise enables the process to move from $y$ to $x$, in spite $x>y$. 
As we did in equation  (\ref{SIeq:ConsistentCont}), we can find a consistency relation for the probability density $p_\lambda$ of visiting a given point of $\Omega$ along a noisy SSRP, 
\begin{equation}
p_\lambda(x)=\lambda\int^N_xp(x|y)p(y)dy+(1-\lambda)q(x).
\label{eq:PConstNoise}
\end{equation}
If we take the derivative
\begin{eqnarray}
\frac{dp_\lambda(x)}{dx}&=&\lambda\frac{d}{dx}\left(\int^N_xp(x|y)p_\lambda(y)dy\right)+(1-\lambda)\frac{dq(x)}{dx}\nonumber\\
&=&\lambda\frac{dq(x)}{dx}\int^N_x\frac{p_\lambda(y)}{g(y)}dy-\lambda\frac{q(x)}{g(x)}p_\lambda(x)+\nonumber\\
&&+
(1-\lambda)\frac{dq(x)}{dx}\nonumber\\
&=&\frac{\lambda}{q(x)}\frac{dq}{dx}\int^N_x\frac{q(x)}{g(y)}p_\lambda(y)dy-\lambda\frac{q(x)}{g(x)}p_\lambda(x)+\nonumber\\
&&+(1-\lambda)\frac{dq(x)}{dx}\nonumber\\
&=&\frac{1}{q(x)}\frac{dq(x)}{dx}(p_\lambda(x)-(1-\lambda)q(x))-\lambda\frac{q(x)}{g(x)}p_\lambda(x)+\nonumber\\
&&+(1-\lambda)\frac{dq(x)}{dx}\nonumber\\
&=&\frac{1}{q(x)}\frac{dq(x)}{dx}p_\lambda(x)-\lambda\frac{q(x)}{g(x)}p_\lambda(x)\quad,\nonumber
\end{eqnarray}
where the fourth step is performed taking the definition of $p_\lambda(x)$ given in equation (\ref{eq:PConstNoise}). 
We therefore have the following differential equation for $p_\lambda(x)$,
\begin{equation}
\frac{dp_\lambda(x)}{dx}=\left(\frac{1}{q(x)}\frac{d q(x)}{d x}-\lambda\frac{q(x)}{g(x)}\right)p_\lambda(x) \quad ,
\end{equation}
which can be rewritten as
\[
\frac{dp_\lambda(x)}{p_\lambda(x)}=\frac{d}{dx}\log\left(\frac{q(x)}{g^\lambda(x)}\right)dx\quad.
\]
Integrating it overall $x\in(1,N]$, we obtain
\begin{eqnarray}
p_\lambda(x)=\frac{1}{Z_\lambda}\frac{q(x)}{g^{\lambda}(x)} \quad,
\label{eq:NoiseContinuous}
\end{eqnarray}
which again demonstrates how the noisy SSRP deforms the underlying prior probabilities $q$,
$Z_\lambda$ being  the normalisation constant. Interestingly, if $\lambda<1$, i.e., if we consider a noisy SSRP, $\lambda$ has the role of a scaling exponent. We observe that
we recover the standard SSRP $\psi$ described above in equation  (\ref{SIeq:P(a_k|a_i)Cont}) if $\lambda\to 1$ (no noise) and the Bernouilli process following the prior probabilities $q$ if we have total noise, as expected. 
The results for the continuous SSRPs are similar to the discrete case; compare equation (\ref{eq:NoiseContinuous}) and equation (\ref{eq:General_Noise}).

\section{Targeted diffusion on networks with different topologies}

In the following we find the mapping between the degree distribution 
$P(k)$ and the undirected ordered degree sequence. Once we know the degree sequence, we can compute the {\em network prior} probabilities $q_G$ thanks to equation  (\ref{eq:PriorNets}). Then, we apply directly equation (\ref{eq:General_Net}), which gives us the general form of statistics of node visits for targeted diffusion.

Without any loss of generality we assume that there is a labelling of the nodes of the graph ${\cal G}$, such that the undirected degree sequence 
$\sigma_G$, given by
\begin{equation}
\sigma_G\equiv\{k_1,. . .,k_N\} \quad ,
\end{equation}
is ordered, meaning that
\begin{equation}
k_1\geq k_2\geq . . .\geq k_N \quad .
\label{eq:ordered}
\end{equation}

In the following we will assume that the degree distribution $P(k)$ is known and that we want to infer the formal shape of $\sigma_G$, if any. In general, a formal mapping from $P(k)$ to $\sigma_G$ is hard or even impossible to find. 
However, it can be approximated. Let us assume that there exists a function $f(i)=k_i$ that 
gives  the degree of the $i$-th node of the ordered degree sequence of the undirected graph ${\cal G}$. 
Suppose, for the sake of notational simplicity, that $k_i=k$. Clearly, 
$f^{-1}(k)=i$. 
From this we infer that there are approximately $i-1$ nodes whose degree is higher than $k$. The probability of finding a randomly chosen node whose degree is higher than $k$, $P_<(k)$, is $P_<(k)=\sum_{k'>k}P(k')$. 
The number of nodes with degree larger than $k$ will thus be approached by $NP_<(k)$. Under the assumption that the number of nodes is large one can argue that 
\begin{equation}
f^{-1}(k) \sim N\int_{k}^{\infty}P(k')dk' \quad .
\label{eq:f^-1}
\end{equation}
The identification 
of $f$ from the knowledge of $P(k)$ provides the functional shape of the ordered degree sequence and, consequently, the {\em network-prior} probability distribution.
\\

{\em Exponential networks:} Exponential networks have a degree distribution given by 
\begin{equation}
P(k)\propto \exp(-\chi k) \quad ,
\end{equation}
with $\chi>0$.
The direct application of equation (\ref{eq:f^-1}) reads
\begin{equation}
f^{-1}(k)\sim N\exp(-\chi k) \quad , 
\end{equation}
leading to 
\begin{equation}
f(i)\sim  \chi^{-1}\log\left(\frac{N}{i}\right) \quad . 
\end{equation}
Since we assumed that $k_i=f(i)$, and knowing, from equation (\ref{eq:PriorNets}), that $q(i)=k_i/2|E|$, the {\em network-prior} probabilities for exponential networks, $q_{\exp}$, are given by
\begin{equation}
q_{\exp}(i)\propto \frac{1}{\chi}\log\left(\frac{N}{i}\right) \quad .
\end{equation}
For large graphs we can approximate $g_G(i)$ by 
\begin{eqnarray}
g_G(i)&=& \sum_{\ell\leq i}q_{\exp}(\ell) \sim \int_1^{i}\log\left(\frac{N}{x}\right)dx\nonumber\\
&\sim& i\log\left(\frac{N}{i}+1\right)+{\cal O}(\log N) \quad ,
\end{eqnarray}
and equation (\ref{eq:General_Net}) asymptotically becomes 
\begin{equation}
p_{\rm }(i)\propto \frac{\log\left(\frac{N}{i}\right)}{i\log\left(\frac{N}{i}+1\right)}\to \frac{1}{i} \quad .
\end{equation}
Targeted diffusion on exponential DAG networks therefore leads to Zipf's law in node visiting frequencies.
\\

{\em Scale-free networks:} Scale-free networks have a degree distribution 
$P(k) \sim k^{-\alpha}$.
For  $\alpha>2$, which is the most common case, one has 
\begin{equation}
f^{-1}(k)\sim Nk^{1-\alpha} \quad ,
\end{equation}
which implies
\begin{equation}
f(i)\sim i^{-\beta} \quad ,
\end{equation}
with $-\beta=(1-\alpha)^{-1}$. Therefore, the {\em network-prior} probabilities for scale-free networks, $q_{SF}$, are given by 
\begin{equation}
q_{SF}(i)\propto i^{-\beta} \quad .
\end{equation}
As a consequence the cumulative {\em network-prior} distribution, $g_{SG}$, is (approximating the sum with an integral) 
\begin{equation}
g_{SF}(i)\sim i^{-\beta+1} \quad .
\end{equation}
Using equation (\ref{eq:General_Net}), this leads to 
\begin{equation}
p_{\rm }(i) \sim \frac{i^{-\beta}}{i^{-\beta+1}}\to \frac{1}{i} \quad .
\end{equation}
Again Zipf's law appears in the node visiting probabilities.

\end{document}